\documentclass[10pt,aps,prl,twocolumn,superscriptaddress]{revtex4-1}
\usepackage{graphicx}
\usepackage{amssymb}
\usepackage{amsmath}
\usepackage{appendix}
\usepackage{comment}

\newcommand{\eq}[1]{\begin{equation}#1\end{equation}}

\usepackage{mdframed}

\begin{document}

\title{Energy Storage via Topological Spin Textures}

\author{Yaroslav Tserkovnyak}
\affiliation{Department of Physics and Astronomy, University of California, Los Angeles, California 90095, USA}
\author{Jiang Xiao}\affiliation{State Key Laboratory of Surface Physics and Department of Physics, Fudan University, Shanghai 200433, China}
\affiliation{Institute for Nanoelectronics Devices and Quantum Computing, Fudan University, Shanghai 200433, China}
\affiliation{Collaborative Innovation Center of Advanced Microstructures, Nanjing 210093, China}

\begin{abstract}
We formulate an energy-storage concept based on the free energy associated with metastable magnetic configurations. Despite the active, magnetic region of the battery being electrically insulating, it can sustain effective hydrodynamics of spin textures, whose conservation law is governed by topology. To illustrate the key physics and potential functionality, we focus here on the simplest quasi-one-dimensional case of planar winding of the magnetic order parameter. The energy is stored in the metastable winding number, which can be injected electrically by an appropriately tailored spin torque. Due to the nonvolatility and the endurance of magnetic systems, the injected energy can be stored essentially indefinitely, with charging/discharging cycles that do not degrade over time.
\end{abstract}

\maketitle

\textit{Introduction.}|The field of spintronics has undergone much progress over the recent years, particularly in regard to memory, logic, and efficient signal propagation \cite{maekawaBOOK17}. Especially intriguing are the magnetic systems based on electrical insulators \cite{soninJETP78,*soninAP10}. Their order-parameter configurations may not only naturally possess metastability and thus memory capabilities (as in the domain-wall \cite{parkinSCI08} and skyrmion \cite{sampaioNATN13} racetracks), but can also transmit information via their collective dynamics \cite{takeiPRL14,*takeiPRB14,lanPRX15,*chumakNATP15}. A particularly robust mode of such transport utilizes topological magnetic textures, such as winding in a quasi-one-dimensional easy-plane (ferro- or antiferro-)magnet \cite{kimPRB15br,*kimPRB16}, skyrmion textures in quasi-two-dimensional magnets \cite{ochoaPRB16,*ochoaPRB17}, or the winding of three-dimensional spin-glass textures \cite{ochoaCM18}. The associated topological ``charge," such as the winding angle or the skyrmion number, is a conserved quantity, whose density obeys a continuity equation and can thus exhibit hydrodynamic behavior. In the easy-plane example, in particular, this hydrodynamics maps onto the problem of a neutral superfluid \cite{halperinPR69,soninJETP78,konigPRL01,takeiPRL14,chenPRB14,*chenPRB14nt}. Importantly, the underlying conservation law is rooted in topology of the order-parameter configuration rather than a symmetry of the associated Lagrangian. Contrary to the more traditional conservation laws, thus, we rely here on the structure of the order-parameter configurations rather than the detailed dynamics. A sizable energy barrier is in practice needed, however, to protect against dissipative processes that can relax the order-parameter texture towards the trivial ground state \cite{halperinIJMPB10,kimPRB15br,ochoaPRB16}.

In this Letter, we show how these dynamic topological spin textures can be exploited for energy storage. A key study case will be provided by metastable spin helices in easy-plane, electrically-insulating magnetic materials. While it has been long known that such textures can store free energy over long times \cite{soninAP10,vedmedenkoPRL14,*dzemiantsovaSR15}, no efficient means for its loading and extraction have been suggested. We propose, to this end, to utilize boundary spin torques produced by adjacent metallic wires \cite{takeiPRL14}. The injected magnetic energy can persist over a long time, in the absence of a load, and can eventually be released to produce useful work in a spintronic circuit by spin-motive forces reciprocal to the spin torque \cite{barnesPRL07,tserkovPRB08mt,tserkovPRB14}. We address several specific realizations of this idea, based on the winding textures in easy-plane materials, skyrmionic magnetic configurations, or spin glasses. We discuss the energy conversion and losses during the battery charging and discharging, stability of the (metastable) storage, and the ultimate upper bound for the achievable energy density in magnetic systems. While at present, it appears challenging to compete with the traditional lithium-ion technology in terms of the energy density, our proposal already has a clear advantage specifically for spintronic circuits, where the topological spin-texture energy storage can be naturally integrated with the nonvolatile logic and memory functionalities \cite{lanPRX15,allwoodSCI05,*khitunJPD10}.

\begin{figure}[pth]
\includegraphics[width=1\linewidth]{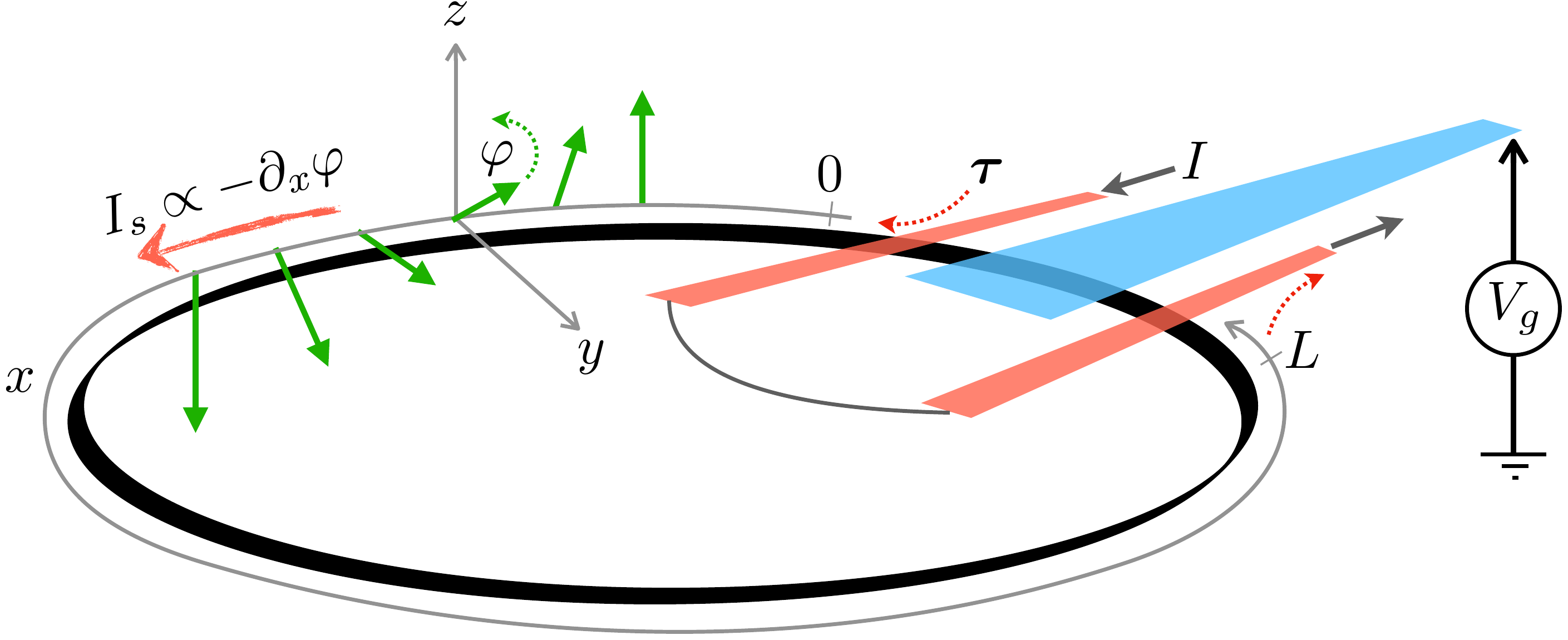}
\caption{A schematic of a quasi-one-dimensional spin spiral of length $L$ that stores energy density $\propto(\partial_x\varphi)^2$, where $\varphi$ is the order-parameter angle within the magnetic easy plane. The winding density $\rho\equiv-\partial_x\varphi$ of the easy-($yz$)-plane directional order parameter is associated with a collective metastable spin current $I_s$ flowing along the loop. This spin current is loaded by applying a spin torque $\boldsymbol{\tau}$ at the ends of the loop, utilizing the spin Hall effect in metallic wires (red) deposited on top. Externally applied electrical current $I$ is thus used to load energy into the battery. This energy is eventually released by exerting a reciprocal spin-motive force onto an external spintronic circuit. The blue gate controls magnetic properties underneath it, allowing to locally interrupt the collective spin current on demand, by turning the gate voltage $V_g$ on/off.}
\label{sch}
\end{figure}

\textit{Prototypical example.}|Fig.~\ref{sch} shows a simplified geometry, which (while not necessarily optimized for a useful device) can be implemented based on existing heterostructures exhibiting spin superfluidity \cite{yuanSA18}. The active spin region forms a loop of length $L$, with the directional order parameter (either of magnetic or N{\'e}el type, for example) constrained to precess within the easy ($yz$) plane. The spins are injected and ejected at the ends of the magnetic wire (i.e., $x=0,L$), where metallic wires can sustain a steady spin torque $\boldsymbol{\tau}$ on the order parameter due to the spin Hall effect \cite{hoffmannIEEEM13,*sinovaRMP15}. The geometry is such that the injected spin current supplied by the left metallic wire (at $x=0$) is polarized along the $x$ axis in spin space. The applied spin current at $x=L$, on the other hand, is polarized in the opposite direction.

In the circuit sketched in Fig.~\ref{sch}, a spin-current flow is, therefore, mediated by the insulating magnetic wire between two metal contacts deposited on top at the two ends. A key requirement is to be able to interrupt this spin current, on demand, in the short region under the top electrostatic gate. This can be achieved by appropriately gating it by the voltage $V_g$, which controls magnetic properties of the magnetic section underneath. When the electric current $I$ is applied to the device (say, during the battery-loading stage), the gate voltage $V_g$ needs to be set to the value that suppresses the local spin current underneath (for example, by reducing the easy-plane anisotropy that is needed to maintain the spin superflow, as discussed below). This can be accomplished according to the effect of the voltage-controlled magnetic anisotropy \cite{weisheitSCI07,*maruyamaNATN09}. When the device is off (in the energy-storage mode), the uniform magnetic properties are restored along the loop, closing a persistent spin superflow, which can then maintain a metastable spin-carrying state over a long time. The battery releases the stored free energy by exerting a spin-motive force on an external circuit \cite{tserkovPRB14}, which is reciprocal to the spin torque that was used to load the battery. For this, we again need to disrupt the persistent superflow by the gate, $V_g$.

Alternatively to the electrostatic gating, local magnetic properties can be controlled by elastic strain \cite{leiNATC13,*finizioPRAP14} or by applying a heat flux to the blue region, which can allow to tune magnetic properties across the magnetic phase transition. The superflow at $x=0^-,L^+$ would certainly be quenched, for example, on the disordered, paramagnetic side of the transition. The variety of the available means to control local magnetic properties makes for versatile battery designs, which may even be integrated or directly endowed with a logic functionality.

\textit{Model.}|Assuming rotational symmetry of the magnetic system around the $x$ axis (in spin space), the $x$ projection of the (nonequilibrium) spin density, $\rho_s(x,t)$, obeys the continuity equation \cite{Note1}
\eq{
\partial_t\rho_s+\partial_x I_s=-\rho_s/\tau_\alpha\,.
\label{ce}}
Expressing the collective spin current \cite{Note2},
\eq{
I_s=-A\partial_x\varphi\,,
\label{Is}}
in terms of the magnetic orientation $\varphi$ within the easy plane (see Fig.~\ref{sch}) is reminiscent of the mass flow in a neutral superfluid. These relations follow from the (phenomenological) magnetic free-energy density (per unit length)
\eq{
F(\rho_s,\partial_x\varphi)=\rho_s^2/2\chi+A(\partial_x\varphi)^2/2\,,
}
parametrized by the spin susceptibility (along the symmetry axis $x$) $\chi$ and the exchange stiffness $A$. Note that we are constructing here a quasi-one-dimensional description, so that the spin density $\rho_s$, the susceptibility $\chi$, and the exchange stiffness $A$ are obtained by multiplying their bulk values by the magnetic wire's cross section.

$\rho_s$, being the generator of the spin order-parameter rotations, is canonically conjugate to $\varphi$ \cite{halperinPR69}:
\eq{
\{\varphi(x),\rho_s(x')\}=\delta(x-x')\,,
}
where the left-hand side stands for the Poisson bracket. The corresponding Hamilton equations lead to the continuity equation \eqref{ce}, once we have supplemented the theory also with the dissipation according to the Rayleigh dissipation function density $R=\alpha s(\partial_t\varphi)^2/2$ \cite{landauBOOKv5}. $\alpha$ is a dimensionless Gilbert damping parameter and $s$ is a convenient normalization constant in units of spin density. (Typically, we choose $s$ to be the full saturation spin density of the underlying spin system.) The spin-relaxation rate on the right-hand side of Eq.~\eqref{ce} is accordingly obtained to be $\tau_\alpha^{-1}=\alpha s/\chi$. The conjugate Hamilton equation is analogous to a local Josephson relation:
\eq{
\partial_t\varphi=\rho_s/\chi\,.
\label{jr}}
The boundary conditions at the metal contacts are constructed according to the spin Hall torque \cite{tserkovPRL17}:
\eq{
-A\partial_x\varphi=\tau=\mp g\partial_t\varphi+\vartheta I\,,
\label{bc}
}
at $x=0,L$. Here, $g\equiv(\hbar/4\pi)g^{\uparrow\downarrow}$, in terms of the interfacial spin-mixing conductance $g^{\uparrow\downarrow}$, and $\vartheta\equiv(\hbar/2e)(S_c/S_w)\tan\theta_{\rm SH}$, in terms of the effective (dimensionless) spin Hall angle $\theta_{\rm SH}$ of the interface. $S_c$  is the spin Hall contact area and $S_w$ is the metallic wire cross section (transverse to the transport direction), so that $I/S_w$ is the applied current density.

\textit{Battery efficiency and capacity.}|For a DC current $I$ applied as sketched in Fig.~\ref{sch} at the two ends of a disconnected magnetic loop, the winding density becomes
\eq{
\rho\equiv-\partial_x\varphi=(\vartheta/A)I\,,
}
spreading uniformly along the magnetic wire, in the steady state. This is determined entirely by the boundary conditions \eqref{bc}. The associated (free) energy stored in the system is
\eq{
E=A\rho^2L/2=(\vartheta I)^2L/2A\,.
}
The time needed to reach the steady state is $t\sim L/u=L\sqrt{\chi/A}$, where $u=\sqrt{A/\chi}$ is the spin-wave speed according to the equations of motion, Eqs.~\eqref{ce} and \eqref{jr}. The associated energy loss due to Joule heating in the metallic wires is thus $Q\sim RI^2t$, where $R$ is the (combined) resistance of the metallic wires or, more generally, the entire circuit supplying the current. The corresponding energy-loss efficiency for ``charging" the battery is
\eq{
\eta=E/Q\sim\vartheta^2/R\sqrt{A\chi}\,.
}
As an example, consider an antiferromagnet, whose $\sqrt{A\chi}\sim\hbar S_m/a^2$ \cite{auerbachBOOK94}, where $a$ is the atomic spacing and $S_m$ is the magnetic wire cross section. Writing $R^{-1}\equiv Ne^2/h$, in terms of the number $N$ of the conductance quanta, we have
\eq{
\eta\sim N\frac{a^2S_c^2}{S_mS_w^2}\tan^2\theta_{\rm SH}\sim\frac{S_c^2}{S_mS_w}\frac{l}{L_w}\tan^2\theta_{\rm SH}\,,
\label{eta}}
having dropped all the numerical factors. In the last approximation, we used $N\sim (S_w/a^2)l/L_w$, which is valid for a long metal wire \cite{beenakkerRMP97}, where $L_w$ is the wire length and $l$ the scattering mean free path.

Geometrically, the efficiency \eqref{eta} benefits from increasing the spin Hall contact area $S_c$ of the charging wire, decreasing the cross section of the magnetic loop $S_m$, and decreasing cross section $S_w$ of the charging wire, while also shortening the charging wire ($L_w$) and making it cleaner (longer mean free path $l$), which are all intuitively clear. So long as the spin Hall angle $\theta_{\rm SH}$ can be of order unity, which is within the reach of the current state of the art \cite{hellmanRMP17}, there is no fundamental reason from preventing a high energy-loss efficiency $\eta$. It may be remarked, however, that when approaching a high efficiency, $\eta\to 1$, the above estimate needs to be refined to properly account for Gilbert damping, spin pumping, as well as dissipative cross terms in the coupled magnetoelectric dynamics, according to the full spin Hall phenomenology \cite{tserkovPRB14}. Our main point here is that the geometry may in principle be optimized (and also improved by going to more complex structures with, e.g., meandering wires for spin injection and detection) in realistic magnetic heterostructures to give a near perfect energy efficiency for loading (as well as, by reciprocity, unloading)  the battery.

Another key concern is the maximal energy storage capacity. The volumetric energy density associated with magnetic winding is
\eq{
\mathcal{E}=\mathcal{A}(\partial_x\varphi)^2/2\,,
}
where $\mathcal{A}\sim J/a$ is the bulk stiffness of the order parameter, in terms of the atomistic exchange $J$. In the extreme limit of the order parameter varying on the atomistic scale, $\partial_x\varphi\sim a^{-1}$. The corresponding maximal energy that can be stored in a magnetic texture is thus of order $J$ per atomic site. Taking the atomic distance of 3~\AA, mass density of a magnetic material of $7\times10^3$~kg/m$^3$, and $J$ of an eV would result in the energy capacity of $\sim200$~W$\cdot$h/kg, which corresponds exactly to the optimal capacity of lithium-ion batteries (that is roughly in the range of an eV per atom). Realistically, however, we may expect to achieve a much lower energy density, as a typical $J$ is an order of magnitude lower in room-temperature ferro- and antiferromagnets, and a more feasible magnetic texture would be one to two orders of magnitude smoother (as explained below). Putting this together, common magnetic materials would yield an energy storage capacity that is lower by at least three orders of magnitude than the currently leading (electrochemical) technology. An additional challenge concerns scaling and compactifying the energy-storying elements so that the magnetic material takes up most of the allotted space, as has been assumed in our estimate.

In order to load the magnetic region with a sharper texture, thus a higher energy density, a current in the metallic wires needs to be raised, according to the boundary conditions \eqref{bc}. However, the magnetic bulk cannot sustain an arbitrarily sharp texture, as eventually a Landau instability will set in, destroying the spiraling texture within the easy plane \cite{soninJETP78}. The associated Landau criterion for the instability is achieved when the winding-texture energy is comparable to the easy-plane anisotropy energy that keeps the winding within a plane, thus endowing it with a topological protection. If we parametrize the latter by $\mathcal{K}$, the critical texture corresponds to $\rho\sim\sqrt{\mathcal{K}/\mathcal{A}}$, which is the inverse of the magnetic healing length for local out-of-the-easy-plane excursions. While we can expect it to be in the range of 10's of nm's (as supposed above), it can reach atomistic scale in extreme cases \cite{heinzeNATP11,*hauptmannNL17}. Ultimately reaching the magnetic energy storage comparable to that of the state-of-the-art electrochemical systems thus appears possible in principle.

One notable potential advantage of the magnetic battery discussed here vs traditional lithium-ion technology is its longevity, as it does not suffer from the ``Coulombic efficiency" of the electron transfer issues that governs the number of charging/discharging cycles within the (electrochemical) battery's life span \cite{smithJES10}. Indeed, magnetic memories based on switching between metastable magnetic states are known for their essentially unlimited endurance \cite{apalkovPIEEE16}, as compared to the electronic counterparts. This bodes well also for the magnetic energy storage. Secondly, the charging speed (ultimately limited by the spin-wave velocity, e.g., $u\sim aJ/\hbar\sim10^5$~m/s in common antiferromagnets) can be faster than the electrochemical processes in a conventional battery. Furthermore, the energy storage discussed here is naturally compatible with spintronic circuits, especially based on magnetic insulators. We can thus envision applications that may simply not be possible with traditional batteries, resulting in low-dissipation operations (including energy storage, logic, and memory) based purely on spin dynamics. Finally, it may be interesting to think of the extensions to (macroscopic) quantum information processing based on magnetic insulators \cite{takeiPRB17}.

\textit{Topological stability.}|The central idea in our proposal is the topological stability of the stored energy. It is analogous to the finite energy density associated with a superfluid state that can maintain its metastable state over years and even decades \cite{halperinIJMPB10}. This is rooted in the topological character of a low-energy theory, here associated with a winding number of the order parameter, and a large energy barrier that needs to be overcome to relax the metastable state via excursions to other topological sectors of the theory. For low-dimensional superfluids and superconductors, such excursions are known as phase slips and can be enacted by thermal or (at very low temperatures) quantum fluctuations \cite{halperinIJMPB10,kimPRB16,kimPRL16}. The above spin-superfluidity based example illustrates the generic aspects of our topological spin textures enabling the (meta)stable energy storage.

Storing energy topologically via winding textures in quasi-one-dimensional systems is not the only possibility that appears potentially accessible to current technology. In two-dimensional magnetic films (both ferro- and antiferromagnetic), a dynamic directional order parameter $\mathbf{n}(x,y,t)$, s.t. $|\mathbf{n}|\equiv1$, defines topological density
\eq{
\rho(x,y,t)=\mathbf{n}\cdot(\partial_x\mathbf{n}\times\partial_y\mathbf{n})\,,
}
which is related to the skyrmion number harbored by the magnetic texture \cite{belavinJETPL75}. This density is globally conserved and locally obeys a two-dimensional continuity equation, which allows one to formulate a topological hydrodynamics \cite{ochoaPRB16}, similarly to that for the winding density \cite{kimPRB15br}. In fact, the topological texture injection can also be carried out in such two-dimensional systems \cite{ochoaPRB16}, albeit using an adiabatic spin-transfer torque \cite{tserkovJMMM08} at the metallic contact instead of the spin Hall effect. Our considerations above, regarding the storage capacity, stability, and efficiency, easily generalize to this case, with the similar final conclusions, both qualitatively and quantitatively. This bodes well, in light of the recent explosive developments of the skyrmionic films \cite{yuNAT10,*huangPRL12,heinzeNATP11} and, particularly, mobile room-temperature skyrmions \cite{jiangSCI15,*jiangAIP16}.

In three spatial dimensions, spin glasses appear promising for enabling topological hydrodynamics and the associated energy storage \cite{tserkovPRB17,wesenbergNATP17,ochoaCM18}. The physics there is analogous to the one-dimensional winding, which gets promoted to three-dimensional winding of the rigid spin-glass arrangements. Injection of the winding textures can be accomplished in spin glasses via the spin Hall effect at metal contacts \cite{tserkovPRB17}. The stability of magnetic textures and the associated maximum energy capacity remain to be important open issues here \cite{ochoaCM18}. In particular, it is interesting whether one can reach a stable energy density at the level of an atomistic exchange per atom, in the case of the \textit{three-dimensional} skyrmionic textures in a spin glass (classified according to the third homotopy group) \cite{Zarzuela}. Indeed, it is the only microscopic energy scale in the case of an isotropic spin glass.

\textit{Summary and discussion.}|It is important to note that the above discussion pertaining to Fig.~\ref{sch} is simplified in that it assumes rotational spin symmetry about the $x$ axis. This is important for the mapping onto superfluid dynamics \cite{takeiPRL14}. While adding anisotropies that would lift this symmetry is quite catastrophic to the underlying \textit{spin} hydrodynamics \cite{soninJETP78,konigPRL01}, it does not conceptually impact our topological energy storage. The central hydrodynamic quantity that can be injected, stored, maintaining a metastable free energy, and eventually unloaded by producing work is the \textit{winding} density $\rho\equiv-\partial_x\varphi$ (or, more generally, an appropriate topological-``charge" density) \cite{kimPRB15br,ochoaPRB16,tserkovPRB17}. In the case of the axial symmetry, the topological, $\rho$, and spin, $\rho_s\propto\partial_t\varphi$, densities obey dual hydrodynamic descriptions, as in the case of a neutral superfluid. While the anisotropies would invalidate pure spin hydrodynamics of $\rho_s$, the topological hydrodynamics of $\rho$ persists, in essence, albeit necessitating some revisions for the quantitative aspects of the theory.

In summary, an energy-storage concept is proposed based on topological hydrodynamics in certain classes of magnetic insulators. While our specific discussions focused on the quasi-one-dimensional easy-plane dynamics (which, in the idealized limit, realizes a spin superfluidity), we also commented on two-dimensional and three-dimensional skyrmionic textures as well as spin glasses, as possible alternative routes towards realizations of topological hydrodynamics. The latter is a required precursor for our energy-storage proposal. While we showed how topologically-stable and energy-carrying magnetic textures can be loaded and unloaded using (thermo)electric means, we ultimately envision their primary utility for feeding integrated spintronic circuits. The magnetic nonvolatility, endurance, high speed and low dissipation of collective sin dynamics, and, ultimately, large energy-storage capacity set by the strong microscopic exchange interactions all bode well for the proposal, warranting further research.

\begin{acknowledgments}
We are grateful to Benedetta Flebus, Weichao Yu, and Yuansheng Zhao for insightful discussions. Y.T. was supported by the U.S. Department of Energy, Office of Basic Energy Sciences under Award No.~DE-SC0012190 and J.X. was supported by the National Natural Science Foundation of China under Grant No.~11722430.
\end{acknowledgments}

\end{document}